\documentclass[9pt, onecolumn]{article}
\usepackage{cite}
\usepackage{amsmath,amssymb,amsfonts}
\usepackage{algorithmic}
\usepackage{graphicx}
\usepackage{textcomp}
\usepackage{lipsum}
\usepackage[table]{xcolor}
\usepackage{nolbreaks}
\usepackage[a4paper, total={6in, 10in}]{geometry}
\def\BibTeX{{\rm B\kern-.05em{\sc i\kern-.025em b}\kern-.08em
    T\kern-.1667em\lower.7ex\hbox{E}\kern-.125emX}}
\begin{document}
% ----------------------------------------------------------------
\title{Efficient Deep Model-Based Optoacoustic Image Reconstruction}
\author{Christoph Dehner, Guillaume Zahnd\\
\textit{iThera Medical GmbH, Munich, Germany}}
\date{Preprint accepted at \\2024 Ultrasonics, Ferroelectrics, and Frequency Control Joint Symposium}
\maketitle
% ----------------------------------------------------------------
\begin{abstract}

Clinical adoption of multispectral optoacoustic tomography necessitates improvements of the image quality available in real-time, as well as a reduction in the scanner financial cost.
Deep learning approaches have recently unlocked the reconstruction of high-quality optoacoustic images in real-time.
However, currently used deep neural network architectures require powerful graphics processing units to infer images at sufficiently high frame-rates, consequently greatly increasing the price tag.
Herein we propose EfficientDeepMB, a relatively lightweight (17M parameters) network architecture achieving high frame-rates on medium-sized graphics cards with no noticeable downgrade in image quality. EfficientDeepMB is built upon DeepMB, a previously established deep learning framework to reconstruct high-quality images in real-time, and upon EfficientNet, a network architectures designed to operate of mobile devices.
We demonstrate the performance of EfficientDeepMB in terms of reconstruction speed and accuracy using a large and diverse dataset of in vivo optoacoustic scans.
EfficientDeepMB is about three to five times faster than DeepMB: deployed on a medium-sized NVIDIA RTX A2000 Ada, EfficientDeepMB reconstructs images at speeds enabling live image feedback (59~Hz) while DeepMB fails to meets the real-time inference threshold (14~Hz).
The quantitative difference between the reconstruction accuracy of EfficientDeepMB and DeepMB is marginal (data residual norms of 0.1560 vs. 0.1487, mean absolute error of 0.642 vs. 0.745).
There are no perceptible qualitative differences between images inferred with the two reconstruction methods.

\end{abstract}
% ----------------------------------------------------------------

% ----------------------------------------------------------------
\section{Introduction}

Multispectral optoacoustic tomography (MSOT) is a non-invasive and non-ionizing functional imaging modality that can detect optical contrast with high spatial resolution and centimeter-scale penetration depth in living tissue~\cite{ntziachristos2010molecular, regensburger2021optoacoustic, dean2022practical, riksen2023photoacoustic, tarvainen2023quantitative, justel2023spotlight}.
Clinical translation of optoacoustic imaging requires both an improvement in the image quality available in real-time~\cite{taruttis2015advances} and a reduction in the scanner financial cost.
In recent research, deep-learning-based image reconstruction methods~\cite{hauptmann2018modelbased, dehner2023deep} have enabled real-time imaging with high image quality.
However, currently used deep neural network architectures (typically full-fledged U-Nets) require powerful graphics cards to infer images in real-time, which significantly adds to the bill of material.
Reducing the computational effort required for image inference would enable financial cost optimizations and advance the clinical translation of optoacoustic tomography.

Herein, we propose a frugal deep convolutional neural network architecture to reconstruct high-quality optoacoustic images in real-time.
We build upon DeepMB~\cite{dehner2023deep}, a previously established deep learning framework, and adapt its deep convolutional neural network layout based on the EfficientNet architecture~\cite{tan2019efficientnet}, which is designed to run on mobile devices with meager computational resources and tight power budgets.
We denote our implementation EfficientDeepMB.

We evaluate EfficientDeepMB in terms of inference time by deploying the network on six different devices with varying computational capabilities, and in terms of image quality with a dataset of 4814 in vivo scans.
EfficientDeepMB enables real-time imaging using a graphics card that is about five times less powerful compared to the one required by DeepMB, with comparable reconstruction accuracy.
% ----------------------------------------------------------------
\section{Methods}

\subsection{Network architecture}

\begin{figure}[h!]
\centering
\includegraphics[width=.58\columnwidth]{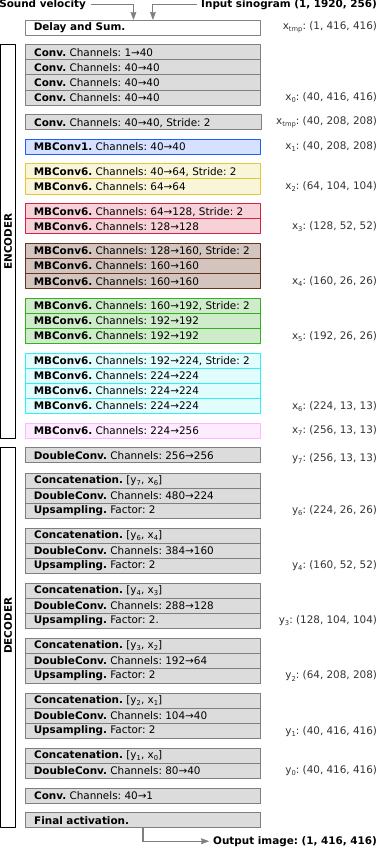}
\caption{
EfficientDeepMB network architecture.
In the encoding pathway, the seven blocks inspired from EfficientNet are shown in color.
The numbers in brackets indicate the tensors shape (channels, height, width).
Conv: block including a 2D convolution, batch normalization, and SiLU activation.
MBConv6: block including an inverted residual block (namely, a pointwise convolution block with expension factor 6, and a depthwise grouped convolution block), a squeeze-and-excitation block with reduction factor 4, a pointwise projection block, and a residual connection block.
MBConv1: similar as MBConv6, albeit without a pointwise convolution block.
DoubleConv: traditional U-Net decoder block, composed of a chain of two Conv blocks.
The size of all convolution kernels is $3\times3$.
Concatenation is applied channel-wise.
}
\label{fig:network}
\end{figure}

Figure~\ref{fig:network} describes the network architecture of \nolbreaks{EfficientDeepMB}.
First, the recorded pressure signals are transformed into the image domain using a delay-and-sum operation (no trainable parameters, and without encoding the speed of sound value as additional channels).
Second, the full-fledged U-Net~\cite{ronneberger2015unet} of DeepMB is replaced by an optimized encoder-decoder-based design of trainable layers:
In the contracting path, an arrangement of inverted residual blocks~\cite{sandler2018mobilenetv2} is used, following the original EfficientNet architecture~\cite{tan2019efficientnet}. 
Inverted residual blocks are composed of a depthwise separable convolution to reduce computational and memory requirements~\cite{chollet2017xception}, a squeeze-and-excitation mechanism to efficiently recalibrate channel-wise feature responses~\cite{hu2018squeeze}, and a residual connection to facilitate training~\cite{he2016deep}.
In the expanding path, the traditional U-Net decoder~\cite{ronneberger2015unet} is employed. 
The original EfficientNet design is adapted by empirically optimizing the scale of the network (in terms of depth, breadth, and resolution) for a compromise between expressiveness and complexity (see Fig.~\ref{fig:network}).

Table~\ref{tab:flops} details the computational cost of the two compared network architectures. The number of computational operations required by EfficientDeepMB is about an order of magnitude lower compared to DeepMB, and the number of learnable parameters is nearly halved.

\begin{table}[h!]
\caption{Comparison of the computational cost of between EfficientDeepMB and DeepMB.
FLOPs: Floating Point Operations.
MACs: Multiply-Accumulate Operations.}
\begin{center}
\begin{tabular}{lcc}
\hline
\rowcolor{brown!15}
& \textbf{EfficientDeepMB} & \textbf{DeepMB} \\
\hline
FLOPs & $52.8 \times 10^9$ & $660.7 \times 10^9$ \\
\rowcolor{brown!7}
MACs & $26.2 \times 10^9$ & $330.0 \times 10^9$ \\
Learnable parameters & $17.4 \times 10^6$ & $32.4 \times 10^6$ \\
\hline
\end{tabular}
\label{tab:flops}
\end{center}
\end{table}

\subsection{Training strategy}

The training strategy used for EfficientDeepMB was the same as for DeepMB~\cite{dehner2023deep}:
Input sinograms were optoacoustic signals synthesized from real-world images from the PASCAL Visual Object Classes Challenge 2012 dataset~\cite{everingham2010pascal}, and target references were optoacoustic images generated by model-based reconstruction~\cite{chowdhury2021individual, chowdhury2020synthetic} of the corresponding signals.
The number of samples in the training dataset and in the validation dataset was 8000 and 2000, respectively.

The EfficientDeepMB network was implemented in Python and PyTorch.
It was trained on synthetic data for 350 epochs using stochastic gradient descent with batch size of 8, learning rate of $1.0\times10^{-2}$, momentum of 0.99, and per-epoch learning rate decay factor of 0.99.
The final activation function was the ReLU function.
The network loss was the smooth L1 loss ($\beta=0.1$) between the predicted image and the reference model-based image.
Gradient norms were clipped to a maximum of 1.0 during backpropagation to prevent spikes in the training loss.
For comparison purposes, a DeepMB network was implemented and trained, with only one modification from the original architecture~\cite{dehner2023deep}: we replaced the mean squared error loss by the smooth L1 loss because we found this improves accuracy.
The two trained PyTorch models were finally compiled into ONNX models for speed-up.

% ----------------------------------------------------------------
\section{Results}

\begin{table*}[t!]
\caption{Comparison of the frame rate (in images per second) between EfficientDeepMB and DeepMB, for graphics cards with different theoretical float32 performance (in trillion floating-point operations per second, TFLOPS).}

\begin{center}
\begin{tabular}{lcccc}
\hline
\rowcolor{brown!15}
& & &  \multicolumn{2}{c}{\textbf{Frame rate}} \\
\rowcolor{brown!15}
& \textbf{Tier} & \textbf{TFLOPS} & \textbf{EfficientDeepMB} & \textbf{DeepMB} \\
\hline
NVIDIA GeForce RTX 4090        & High-end & 82.6 & 182.3 & 68.5  \\
\rowcolor{brown!7}
NVIDIA GeForce RTX 3090        & High-end & 35.6 & 108.9 & 30.3 \\
NVIDIA RTX A2000 Ada           & Medium   & 12.0 & 50.9  & 14.3 \\
\rowcolor{brown!7}
NVIDIA GeForce RTX 2060 SUPER  & Medium   & 7.2  & 42.1 & 10.4 \\
NVIDIA GeForce RTX 3060 Mobile & Mobile   & 10.9 & 40.7 & 10.0 \\
\rowcolor{brown!7}
NVIDIA Jetson Xavier AGX       & Mobile   & 1.4  & 6.1 & 1.4 \\
\hline
\end{tabular}
\label{tab:gpus_and_framerate}
\end{center}
\end{table*}

\begin{figure}[t!]
\centering
\includegraphics[width=0.58\columnwidth]{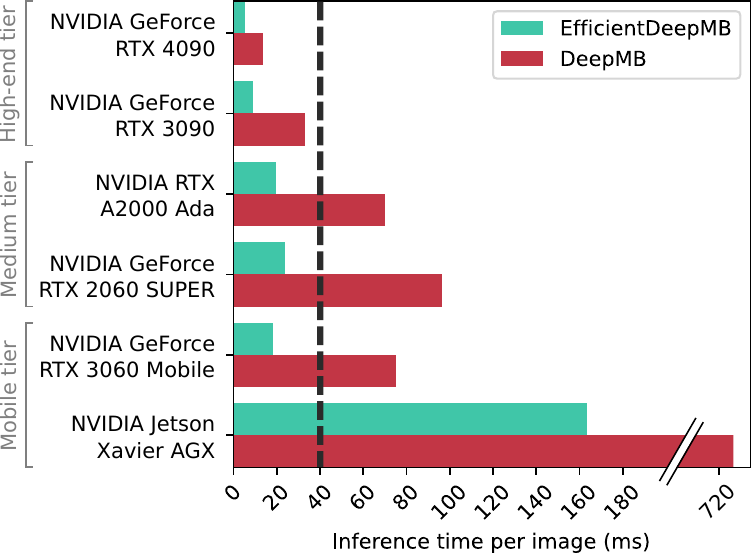}
\caption{Comparison of the inference time between EfficientDeepMB and DeepMB, for different graphics cards.
The dashed line represents the threshold for real-time imaging.}
\label{fig:timing}
\end{figure}

\subsection{Reconstruction speed}

To demonstrate the performance of EfficientDeepMB in terms of inference time, we deployed the compiled models on six different devices equipped with graphics cards of varying computational power, as shown in Table~\ref{tab:gpus_and_framerate}.
Figure~\ref{fig:timing} compares the average end-to-end inference time between EfficientDeepMB and DeepMB for all the considered devices.
While both methods are real-time capable on the most powerful graphics cards (see Fig.~\ref{fig:timing}, high-end tier), only EfficientDeepMB achieves a frame rate suitable for real-time imaging on graphics cards with moderate computational power (see Fig.~\ref{fig:timing}, medium tier).
The two bottom rows (see Fig.~\ref{fig:timing}, mobile tier) demonstrate that EfficientDeepMB can operate live on a laptop, and hints towards the potential for further EfficientDeepMB-enabled miniaturization on an embedded computing board.

All corresponding frame rate values are given in Table~\ref{tab:gpus_and_framerate}.
For comparison, the model-based reference reconstruction algorithm requires 30--60 seconds per image on the high-end GPU and is therefore prohibitive for real-time imaging.

\subsection{Reconstruction accuracy}

To evaluate the capability of EfficientDeepMB to reconstruct high-quality images, we used the in vivo dataset from the original DeepMB study~\cite{dehner2023deep} (4814 scans, six participants, up to eight anatomical regions per participant), acquired with a modern hand-held optoacoustic scanner (MSOT Acuity Echo, iThera Medical GmbH).

\begin{figure*}[b!]
\centering
\includegraphics[width=1\textwidth]{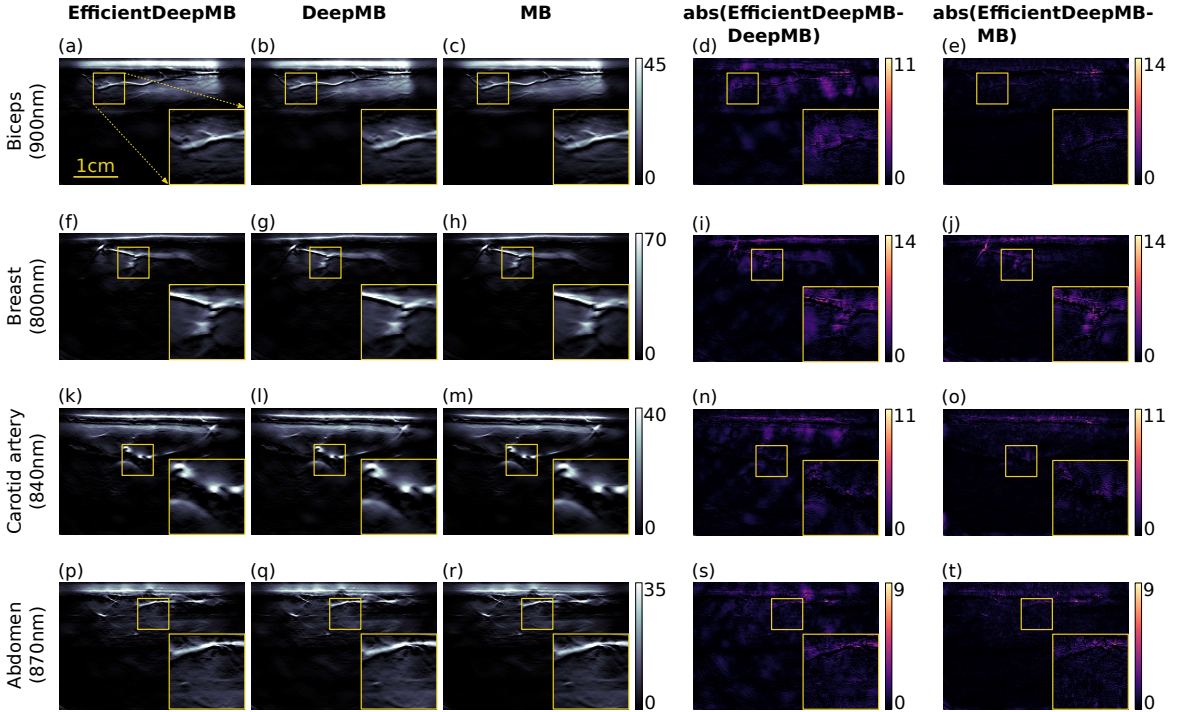}
\caption{Representative examples of optoacoustic images from the in vivo test dataset for different anatomical locations, reconstructed with \nolbreaks{EfficientDeepMB} (a,~f,~k,~p), DeepMB (b,~g,~l,~q), and model-based (MB) (c,~h,~m,~r).
The last two columns show the mean absolute difference between EfficientDeepMB and DeepMB (d,~i,~n,~s), as well as between EfficientDeepMB and MB (e,~j,~o,~t).
For each row, the value within brackets indicates the laser wavelength.}
\label{fig:reconstructions}
\end{figure*}

Figure~\ref{fig:reconstructions} displays example images reconstructed from four different in vivo scans.
This qualitative evaluation shows that EfficientDeepMB images (Fig.~\ref{fig:reconstructions}a,~f,~k,~p) are nearly indistinguishable from both their DeepMB counterparts (Fig.~\ref{fig:reconstructions}b,~g,~l,~q) and the target model-based references (Fig.~\ref{fig:reconstructions}c,~h,~m,~r).
A careful visual examination of all 4814 reconstructed samples of the test dataset confirmed that there were no perceptible differences between the three reconstruction methods, and verified the absence of any noticeable failures, outliers, or artefacts.

Table~\ref{tab:metrics} presents a qualitative evaluation of the reconstruction accuracy for all 4814 samples of the test dataset.
Data residual norms measure the fidelity of the reconstruction process.
Data residual norms of EfficientDeepMB are almost as low as data residual norms of the reference model-based algorithm, and comparable to the data residual norms of DeepMB.
The other metrics (mean absolute error, relative man absolute error, mean squared error, relative mean squared error, peak signal-to-noise ratio, structural similarity index) measure the similarity of the inferred images against model-based reconstructions, and attest that EfficientDeepMB is similarly accurate as DeepMB.

\begin{table*}[b!]
\caption{
Quantitative evaluation of the image quality for EfficientDeepMB and DeepMB, compared against reference model-based (MB) reconstructions. 
The table shows the mean values and in brackets the 25$^\text{th}$ and 75$^\text{th}$ percentiles for all 4814 images of the in vivo test dataset.
The arrow symbols ($\uparrow$ and $\downarrow$) indicate for each metric whether a higher or lower value is better. 
R, data residual norm; MAE, mean absolute error; MAE$_\text{rel}$, relative man absolute error; MSE, mean squared error; MSE$_\text{rel}$, relative mean squared error; PSNR: peak signal-to-noise ratio; SSIM, structural similarity index.
}
\begin{center}
\begin{tabular}{lccc}
\hline
\rowcolor{brown!15}
& \textbf{EfficientDeepMB} & \textbf{DeepMB} & \textbf{MB} \\
\hline
R ($\downarrow$)                    & 0.1560 & 0.1487  & 0.1411 \\
& {\scriptsize (0.0881, 0.1938)}    & {\scriptsize (0.0818, 0.1849)} & {\scriptsize (0.0694, 0.1839)} \\ 
\rowcolor{brown!7}
MAE ($\downarrow$)                  & 0.642 & 0.745  & - \\
\rowcolor{brown!7}
& {\scriptsize (0.358, 0.626)}    & {\scriptsize (0.465, 0.770)} & \\ 
MAE$_\text{rel}$ (\%, $\downarrow$) & 12.79 & 15.65  & - \\
& {\scriptsize (10.63, 14.47)}    & {\scriptsize (14.07, 17.22)} & \\ 
\rowcolor{brown!7}
MSE ($\downarrow$)                  & 6.902 & 5.975  & - \\
\rowcolor{brown!7}
& {\scriptsize (0.429, 1.691)}    & {\scriptsize (0.581, 2.190)} & \\ 
MSE$_\text{rel}$ (\%, $\downarrow$) & 1.02 & 1.14  & - \\
& {\scriptsize (0.50, 1.13)}    & {\scriptsize (0.74, 1.31)} & \\ 
\rowcolor{brown!7}
PSNR (dB, $\uparrow$)               & 46.01 & 44.99  & - \\
\rowcolor{brown!7}
& {\scriptsize (44.39, 47.49)}    & {\scriptsize (43.28, 46.53)} & \\ 
SSIM ($\uparrow$)                   & 0.99 & 0.98  & - \\
& {\scriptsize (0.98, 0.99)}    & {\scriptsize (0.97, 0.99)} & \\ 
\hline
\end{tabular}
\label{tab:metrics}
\end{center}
\vspace{50pt}
\end{table*}

% ----------------------------------------------------------------
\section{Conclusion}

We propose EfficientDeepMB, a frugal deep neural network architecture capable of reconstructing high-quality optoacoustic images in real-time when deployed on medium-sized graphics processing units.
Compared against DeepMB, a recently introduced deep learning framework, EfficientDeepMB can infer images at speeds enabling live image feedback using devices about five times less powerful, with no downgrade in reconstruction accuracy.
EfficientDeepMB paves the way towards miniaturization of the MSOT technology and clinical translation of the modality.

% ----------------------------------------------------------------

% ----------------------------------------------------------------

\end{document}